\documentclass[aps, prb, amsfonts, amssymb, reprint, twoside, superscriptaddress]{revtex4-2}
\usepackage[english]{babel}
\usepackage[utf8]{inputenc}
\usepackage[colorinlistoftodos, color=green!40, prependcaption]{todonotes}
\usepackage{xcolor}
\usepackage{csquotes}
\usepackage{siunitx}
\usepackage{textcomp}
\usepackage{gensymb}
\usepackage{array}
\usepackage{longtable}
\usepackage{physics}
\usepackage{braket}
\usepackage{enumitem}
\usepackage[pdftex, pdftitle={Article}, pdfauthor={Author}]{hyperref}
\usepackage[capitalise]{cleveref}

\newcommand{\siz}{\hat{\sigma}_i^z}
\newcommand{\sjz}{\hat{\sigma}_j^z}
\newcommand{\six}{\hat{\sigma}_i^x}
\newcommand{\sjx}{\hat{\sigma}_j^x}
\newcommand{\siy}{\hat{\sigma}_i^y}
\newcommand{\sjy}{\hat{\sigma}_j^y}
\newcommand{\Wi}{\hat{W}_i}
\newcommand{\Vj}{\hat{V}_j}

\renewcommand{\selectlanguage}[1]{}

\begin{document}
\let\flushbottom\relax
\raggedbottom

\title{Quantum information scrambling in strongly disordered Rydberg spin systems}

\author{Maximilian Müllenbach}
\thanks{These authors contributed equally to this work.}
\affiliation{Centre Européen des Sciences Quantiques (CESQ-ISIS, UMR7006), Université de Strasbourg et CNRS, 23 Rue du Loess, 67200 Strasbourg, France}
\affiliation{Physikalisches Institut, Universit\"at Heidelberg, Im Neuenheimer Feld 226, 69120 Heidelberg, Germany}
\author{Sebastian Geier}
\thanks{These authors contributed equally to this work.}
\affiliation{Physikalisches Institut, Universit\"at Heidelberg, Im Neuenheimer Feld 226, 69120 Heidelberg, Germany}
\author{Adrian Braemer}
\affiliation{Physikalisches Institut, Universit\"at Heidelberg, Im Neuenheimer Feld 226, 69120 Heidelberg, Germany}
\author{Eduard J. Braun}
\affiliation{Physikalisches Institut, Universit\"at Heidelberg, Im Neuenheimer Feld 226, 69120 Heidelberg, Germany}
\author{Titus Franz}
\affiliation{Physikalisches Institut, Universit\"at Heidelberg, Im Neuenheimer Feld 226, 69120 Heidelberg, Germany}
\author{Gerhard Zürn}
\affiliation{Physikalisches Institut, Universit\"at Heidelberg, Im Neuenheimer Feld 226, 69120 Heidelberg, Germany}
\author{Matthias Weidemüller}
\affiliation{Physikalisches Institut, Universit\"at Heidelberg, Im Neuenheimer Feld 226, 69120 Heidelberg, Germany}
\author{Martin Gärttner}
\thanks{Contact author: martin.gaerttner@uni-jena.de}
\affiliation{Institut für Festkörpertheorie und -optik, Friedrich-Schiller-Universität Jena, Max-Wien-Platz 1, 07743 Jena, Germany}

\date{\today} 

\begin{abstract}
Despite the fact that power-law interactions occur in a plethora of physical systems, their many-body dynamics is far less understood than that of nearest-neighbor interacting systems. Here, we study information scrambling in strongly disordered spin systems with power-law interactions via out-of-time-order correlators (OTOCs). Numerically, we find pronounced differences in the dynamical spreading of OTOCs between nearest-neighbor and power-law interacting systems. This deviation persists even for short-range interactions, opposing the common view that these interactions produce dynamics equivalent to the nearest-neighbor case. In a detailed experimental proposal, tailored but not limited to Rydberg tweezer setups, we present a protocol to extract OTOCs in XXZ Heisenberg spin systems with tunable anisotropy and programmable disorder based on currently available techniques.
\end{abstract}

\maketitle

\section{Introduction}
The dynamical evolution of information in quantum many-body systems is the key to understanding many physical phenomena. Questions such as how fast correlations can build up, how hard a system is to simulate classically, and whether thermalization occurs, are intricately linked to the way quantum information spreads \cite{lewis-swan_dynamics_2019}. Although the total information entropy is conserved under unitary evolution, information is distributed throughout the system and becomes inaccessible to local probes during the interaction dynamics. This process called scrambling can be captured by the growth of initially local operators \cite{swingle_unscrambling_2018,xu_scrambling_2024}. In the Heisenberg picture, the spatiotemporal evolution of an operator $\hat{A}_i$, acting only on site $i$ at $t=0$, can be probed by measuring the commutator
\begin{align}\label{eq:OTO-commuator}
    C(r,t) = \norm{\left[\hat{A}_i(t), \hat{B}_{i+r}(0)\right]}_F^2,
\end{align}
where $\hat{B}_{i+r}$ acts on site $i+r$. Here $\norm{\cdot}_F^2$ denotes the Frobenius norm, and due to its close connection with the out-of-time-order correlator (OTOC)  \cite{swingle_unscrambling_2018,xu_scrambling_2024,lieb_finite_1972,luitz_information_2017} $\left\langle\hat{A}_i(t)\hat{B}_{i+r}(0)\hat{A}_i(t)\hat{B}_{i+r}(0)\right\rangle$, $C(r,t)$ is called the out-of-time-order commutator. While there is no universal speed limit for the propagation of information in non-relativistic quantum mechanics, its spread is bounded in short-range interacting systems. Lieb and Robinson (LR) have proven that for finite-range interactions, $C(r,t)$ decays exponentially outside a linear ``light cone'' $t>r/v$, where $v$ is a system-dependent velocity \cite{lieb_finite_1972,nachtergaele_much_2011,cheneau_light-cone-like_2012,chen_speed_2023}. The LR theorem limits the formation of correlations to a causal region and recently found applications in spectral theory, gate complexity, and in bounding finite-size errors \cite{hastings_lieb-schultz-mattis_2004,bravyi_lieb-robinson_2006,hastings_area_2007,hastings_quantization_2015,haah_quantum_2021,wang_bounding_2021}. In disordered systems, the spread of quantum information is slowed down and $C(r,t)$ shows features allowing one to discriminate between Anderson localization \cite{anderson_absence_1958} and the strongly disordered interacting regime associated with many-body localization \cite{abanin_colloquium_2019}. In the latter, the effective $\hat{\tau}^z_i\hat{\tau}^z_j$ interactions of exponentially localized charges induce dephasing and, at least at finite times and system sizes, lead to logarithmic entanglement growth and logarithmic light cones $\ln t \gtrsim r$, whereas information spreading is arrested in Anderson localization \cite{burrell_bounds_2007,kim_local_2014,chen_universal_2016,chen_out--time-order_2017,swingle_slow_2017,fan_out--time-order_2017,huang_out--time-ordered_2017,he_characterizing_2017,bordia_out--time-ordered_2018,lee_typical_2019,kim_slowest_2023}.

Experimental quantum systems, such as Rydberg atoms, trapped ions, NV centers, or nuclear spins in solids (NMR) exhibit long-range power-law interactions $\propto1/r^\alpha$, which can violate the LR bound. Faster than linear information spreading has been observed, both in theory \cite{hastings_spectral_2006,eisert_breakdown_2013,hauke_spread_2013,damanik_new_2014,foss-feig_nearly_2015} and in experiments \cite{richerme_non-local_2014,jurcevic_quasiparticle_2014}, for $\alpha<2d+1$, with $d$ being the spatial dimension of the system under study. The long search for tight and rigorous bounds in power-law interacting systems has only recently been resolved \cite{chen_finite_2019,kuwahara_strictly_2020,tran_hierarchy_2020,tran_lieb-robinson_2021}. Tight bounds $t\gtrsim r^{\min(\alpha-2d,1)}$ have been found for $2d<\alpha<2d+1$, and for $\alpha>2d+1$, linear light cones have been shown to always be an upper bound \cite{tran_lieb-robinson_2021,kuwahara_strictly_2020}. Moreover, the Frobenius norm restricts information spreading to linear light cones already at $\alpha>5/2$ (in one dimension) for the vast majority of initial states $\ket{\psi}$ \cite{tran_hierarchy_2020}. Using quantum typicality and Krylov methods, Luitz and Colmenarez have investigated OTOCs in large ordered Heisenberg systems with power-law interactions \cite{luitz_emergent_2019,colmenarez_lieb-robinson_2020}. Although they established different short-time scalings $C^{pl}(r,t)\propto t^2/r^{2\alpha}$ for power-law vs. $C^{nn}(r,t)\propto (t^r/r!)^2$ for nearest-neighbor interactions, these can be seen as corrections with magnitudes less than $10^{-4}$ for dipolar interactions, and the power-law curves quickly merge with the nearest-neighbor results as $Jt$ increases. These recent advances further support the notion that power-law interactions are essentially nearest-neighbor up to small corrections if $\alpha$ is large enough.

In the first part of this work, we show that this expectation fails for strongly disordered systems. To this end, we investigate operator growth in the Heisenberg XXZ model with disordered on-site potentials. While for nearest-neighbor interactions, the system exhibits well-established logarithmic light cones $\ln t\gtrsim r$ \cite{burrell_bounds_2007,kim_local_2014,chen_universal_2016,chen_out--time-order_2017,swingle_slow_2017,fan_out--time-order_2017,huang_out--time-ordered_2017,he_characterizing_2017,bordia_out--time-ordered_2018,lee_typical_2019,kim_slowest_2023}, we numerically find distinctive algebraic light cones for both dipolar ($\alpha=3$) and van der Waals ($\alpha=6$) interactions. Identifying the underlying mechanism, we conjecture that this result holds for even larger $\alpha$ for sufficiently strong disorder. 

In the second part of this work, we present a protocol for measuring OTOCs in Rydberg atom arrays, implementing Heisenberg XXZ models with tunable anisotropies and on-site disorder \cite{browaeys_many-body_2020,geier_floquet_2021,scholl_microwave_2022,steinert_spatially_2023,franz_observation_2024,bornet_enhancing_2024}. One prominent method to extract OTOCs, complementary to randomized measurements \cite{vermersch_probing_2019,nie_detecting_2019,joshi_quantum_2020}, are echo experiments involving backward evolution $\hat{U}^\dagger = e^{i\hat{H}t}$ \cite{garttner_measuring_2017,wei_exploring_2018,wei_emergent_2019,sanchez_perturbation_2020,nie_experimental_2020,pegahan_energy-resolved_2021,mi_et_al_information_2021,braumuller_probing_2022,zhao_probing_2022,xiang_observation_2024,liang_observation_2025,abanin_et_al_observation_2025}. However, implementing this time-reversal poses a significant experimental challenge. We propose an efficient Floquet driving that reverses both interactions and disordered potentials. Finally, we investigate different strategies for estimating the relevant infinite temperature expectation values of OTOCs through initial state sampling and compare them under practical experimental considerations.

\section{Operator growth in the strongly disordered regime\label{sec:theory}}
\subsection{Model and simulation method}
We consider the Heisenberg XXZ model of $N$ interacting spin-1/2 particles on a one-dimensional chain with random on-site potentials.
\begin{align}
    \label{eq:model_ordered}
    \hat{H} = \frac{1}{2}\sum_{i,j}{J_{ij}\left(\six\sjx+\siy\sjy+\Delta\siz\sjz\right)}+\sum_i{h_i\siz}
\end{align}
The coefficients $h_i$ are uniformly drawn from $[-h,h]$, where $h$ denotes the disorder strength. Motivated by NMR spin echo experiments \cite{wei_exploring_2018,wei_emergent_2019}, the anisotropy parameter is chosen to be $\Delta = -2$. Interactions $J_{ij}$ are either nearest-neighbor or power-law $J_{ij}=C_{\alpha}/r^{\alpha}$. Power-law interactions naturally arise in neutral atom arrays, where the interaction range in the resulting effective spin models depends on the chosen atomic levels \cite{browaeys_many-body_2020, geier_floquet_2021, scholl_microwave_2022, steinert_spatially_2023, franz_observation_2024, morgado_quantum_2021}. While van der Waals interactions ($\alpha=6$) typically arise when encoding a spin in an atomic ground state and a Rydberg state or different $\ket{nS}$ and $\ket{n'S}$ Rydberg states, dipolar interactions ($\alpha=3$) are found as interactions between $\ket{nS}$ and $\ket{n'P}$ Rydberg states. In accordance with the proposed experimental realization in \cref{sec:exp}, we focus on dipolar interactions ($\alpha=3$) but also extend our observations to vdW-interactions further below. In the case of nearest-neighbor couplings, the Heisenberg XXZ model has been reported to undergo a finite-size ergodic to non-ergodic transition at $h_c/|J_z| \approx 3.5$, i.e.,~$h_c\approx7$, associated with many-body localization \cite{znidaric_many-body_2008,pal_many-body_2010,luitz_many-body_2015,abanin_recent_2017,abanin_colloquium_2019}\footnote{Whether this transition persists as a genuine dynamical phase transition in the thermodynamic limit is currently under debate \cite{suntajs_quantum_2020,morningstar_avalanches_2022, abanin_distinguishing_2021, sierant_many-body_2025}. Throughout this work, the \enquote{strongly disordered regime} refers to the operationally defined regime of finite systems and times accessible to numerics and quantum simulation experiments. Our conclusions do not rely on the fate of many-body localization in the thermodynamic limit.}. We investigate operator growth in the strongly disordered regime ($h=14$) via the Frobenius norm of out-of-time-order commutators of two Pauli-$x$ operators at distance $r=|i-j|$
\begin{align}
    \nonumber
    C_x(r,t) &= \norm{[\six(t),\sjx]}_F^2 \\
    \nonumber
    &= \frac{1}{2^N} \Tr\left([\six(t),\sjx][\six(t),\sjx]^\dagger\right)\\
    &= 2 \left[1-\Re\left(\frac{1}{2^N}\Tr\left(\six(t)\sjx\six(t)\sjx\right)\right)\right].\label{eq:xx-OTOC}
\end{align}
We begin by comparing operator growth for nearest-neighbor and dipolar interactions in the strongly-disordered regime. To maximize attainable distances $r = j-i$, we work with open boundary conditions and choose $i = 3$ to avoid influences from reflections at the boundary. The Frobenius norm corresponds to the infinite temperature expectation value and yields the average or \textit{typical} modes of operator growth, while the operator norm, used in the original LR theorem, relates to the fastest mode \cite{colmenarez_lieb-robinson_2020}. We focus on the OTO commutator of two Pauli-$x$ operators since it is not reducible to two-time correlators as $\six$ has zero overlap with the conserved charge $\hat{S}^z=\sum_i{\siz}$ of the Heisenberg XXZ model \cite{balachandran_slow_2023}. There are two computationally costly operations in evaluating \cref{eq:xx-OTOC}: the time evolution and the many matrix-matrix multiplications. We use exact diagonalization to compute the time evolution of $\six(t)$ in the Heisenberg picture. To speed up the evaluation of \cref{eq:xx-OTOC}, we rely on the concept of quantum typicality \cite{steinigeweg_typicality_2016,luitz_information_2017,luitz_emergent_2019,colmenarez_lieb-robinson_2020}, which is a consequence of Lévy's lemma on concentration of measure \cite{levy_addition_1939,milman_asymptotic_2001,heitmann_selected_2020}. In high-dimensional Hilbert spaces, infinite-temperature expectation values $1/2^N\Tr\hat{A}$ are well approximated by $\bra{\psi}\hat{A}\ket{\psi}$, where $\ket{\psi}$ is a \textit{typical} state, i.e., drawn from the Haar measure. Substituting the trace in \cref{eq:xx-OTOC} by $\bra{\psi}\six(t)\sjx\six(t)\sjx\ket{\psi}$, we can compute $C_x(r,t)$ by matrix-vector multiplications instead. The error in the approximation decreases exponentially in system size and, depending on system size, averaging over as few as 1 - 100 initial Haar random states (cf. \cref{subsec:sampling,app:typicality}) gives accurate results. For the system size $N=13$ used here, we choose to average over 10 states per disorder realization.

\subsection{Algebraic Lightcones}
\begin{figure}[tbp]
    \centering
    \includegraphics[scale=1]{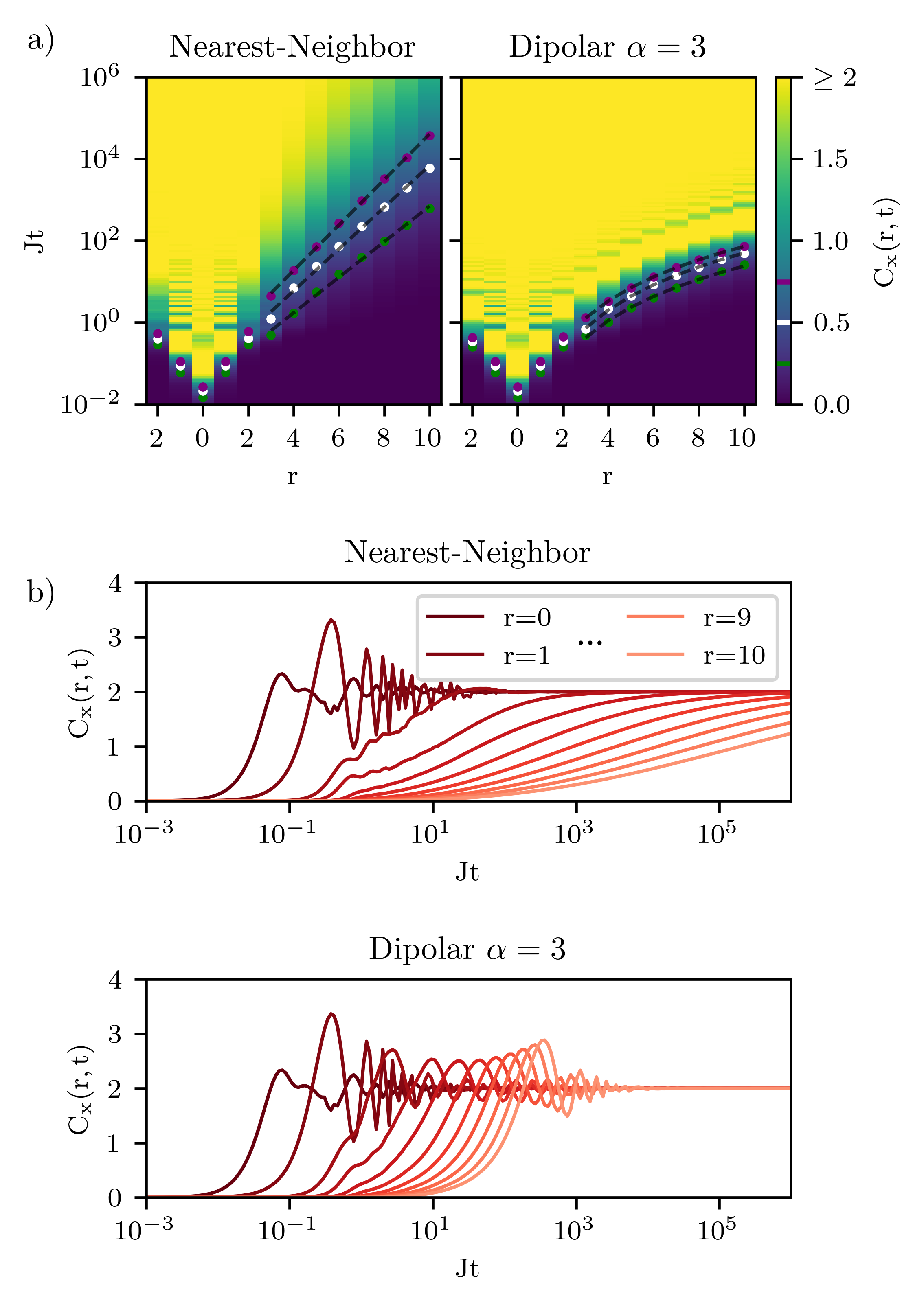}
    \caption{\textbf{Operator growth in the strongly disordered regime.} \textbf{a)} Out-of-time-order commutators $C_x(r,t)=\norm{[\hat{\sigma}^x_3(t),\sjx]}_F^2$ for the Heisenberg XXZ model with nearest-neighbor or dipolar interactions at strong disorder $h=14$. Shown is the disorder average over 5000 individual disorder realizations at a system size of $N=13$. The indicated marks correspond to thresholds $C_x(r,t_\theta)=\theta$ with $\theta \in \{0.25,\,0.5,\,1\}$ and dashed lines are fits $t_\theta \propto e^{\beta r}$, and $t_\theta \propto r^\beta$. \textbf{b)} Vertical cuts through the light cones in (a) showing the time evolution of $C_x(r,t)$ at different distances $r$.}
    \label{fig:lightcones}
\end{figure}
At very strong disorder ($h=14$), the long-time dynamics of $C_x(r,t)$ differ substantially between power-law and nearest-neighbor interactions. \Cref{fig:lightcones}~a) showcases the disorder-averaged operator growth $C_x(r,t)$ for both interactions as functions of distance $r$ and (normalized) time $Jt$, yielding characteristic \textit{light cones}. For nearest-neighbor interactions, we recover the established logarithmic light cones $\ln t \sim r$ (note the logarithmic axis for $Jt$) for operator growth in the many-body localized regime \cite{burrell_bounds_2007,kim_local_2014,chen_universal_2016,chen_out--time-order_2017,swingle_slow_2017,fan_out--time-order_2017,huang_out--time-ordered_2017,he_characterizing_2017,bordia_out--time-ordered_2018,lee_typical_2019,kim_slowest_2023}. For dipolar interactions, we see that the operators grow much faster. The contour lines $C_x(r,t_\theta) = \theta$ follow power-law curves $t_\theta \propto r^\beta$. Fitting the tail ($r>2$) for different thresholds $\theta$, we find $\beta = 1.478 \pm 0.041$, suggesting $\beta\approx\alpha/2$. However, this suggested scaling does not stand up to scrutiny as $\beta/\alpha$ varies with threshold $\theta$ and disorder strength $h$. \Cref{fig:lightcones}~b) shows $C_x(r,t)$ as functions of time for different distances $r>0$, i.e., vertical cuts through the light cones above in \cref{fig:lightcones}~a). For both types of interaction, we see different phases of growth. $C_x(r,t)$ first grows rapidly, which is especially visible for nearby sites and is consistent with previous short-time calculations \cite{luitz_emergent_2019,colmenarez_lieb-robinson_2020}. For nearest-neighbor interactions and $r\geq2$, $C_x(r,t)$ grows more slowly and takes several orders of magnitude longer to reach the saturation value of $C_x=2$, where OTOCs $\left\langle\six(t)\sjx\six(t)\sjx\right\rangle$ vanish. In the case of dipolar interactions and $r\geq2$, we find faster operator growth in the second phase and stronger oscillations around $C_x=2$, which eventually fade out.

\begin{figure}[!ht]
    \centering
    \includegraphics[scale=1]{figs/distributions_obc.png}
    \caption{\textbf{Ensemble of Disorder Realizations} Distribution of out-of-time-order commutators $C_x(r,t)$ over 5000 individual disorder realizations ($h=14$) for nearest-neighbor and dipolar interactions, exemplified for $r=3$ ($N=13$). The red curves correspond to the disorder average and the probability of $C_x(r=3,t)$ to take a given value is indicated by the estimated probability density functions (PDF, color background). The dashed white curve in the lower panel shows the (disorder-independent) analytical solution $C^I_x(r=3,t)$ for power-law Ising interactions.}
    \label{fig:distributions}
\end{figure}

\begin{figure*}[!ht]
    \centering
    \includegraphics[width=\textwidth]{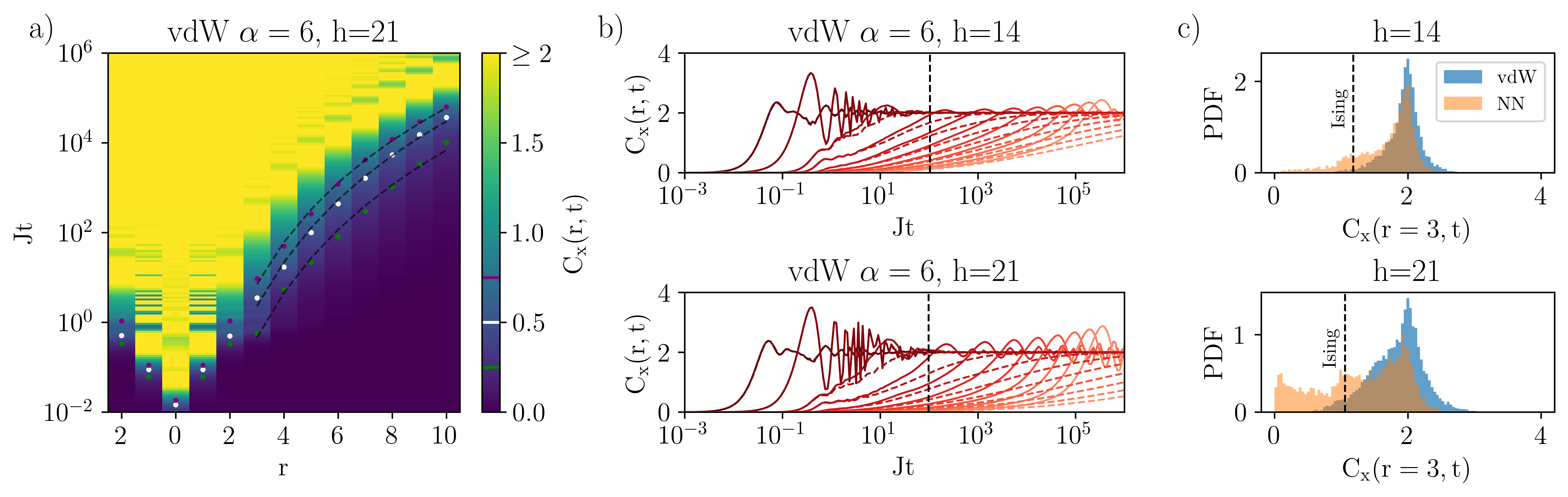}
    \caption{\textbf{Operator Growth for vdW interactions} \textbf{a)} Out-of-time-order commutators $C_x(r,t)=\norm{[\hat{\sigma}^x_3(t),\sjx]}_F^2$ for the Heisenberg XXZ model with vdW interactions at strong disorder $h=21$. Shown is the disorder average over 5000 individual disorder realizations at a system size of $N=13$. The indicated marks correspond to thresholds $C_x(r,t_\theta)=\theta$ with $\theta \in \{0.25,\,0.5,\,0.75\}$ and dashed lines are fits $t_\theta \propto r^\beta$. \textbf{b)} Time evolutions of $C_x(r,t)$ at different distances $r$ for nearest-neighbor (dashed) and vdW (solid) interactions at $h=14$ and $h=21$. The distribution of $C_x(r=3,t)$ sampled across 5000 disorder realizations at $Jt\sim 100$ (vertical dashed line) is illustrated in \textbf{c)}. The vertical line indicates the value of $C^I_x$ and the fractions of realizations with slower growth are $P_{\mathrm{vdW}}=0.02$, $P_{\mathrm{NN}}=0.15$ at $h=14$ and $P_{\mathrm{vdW}}=0.04$, $P_{\mathrm{NN}}=0.34$ at $h=21$.}
    \label{fig:vdW}
\end{figure*}

To gain a more detailed understanding of the differences between power-law and nearest-neighbor interaction dynamics, it is instructive to investigate the ensemble of individual disorder realizations. In fact, individual disorder realizations feature oscillations for nearest-neighbor interactions, but shot-to-shot they are out of phase and average out. In \Cref{fig:distributions}, we depict the ensemble of 5000 disorder realizations by plotting the probabilities for $C_x(r=3,t)$ to assume a given value at time $Jt$. For both types of interaction, we can observe that the peak of the distribution shifts from $C_x\approx0$ to $C_x\approx 2$ at late times. As previously noted by Lee et al. \cite{lee_typical_2019} for nearest-neighbor interactions, the transfer is not described by a moving unimodal distribution, but shows multimodal behavior. Thus, the disorder average alone fails to fully portray operator growth in the strongly disordered regime. The left edge of the initial growth phase consists of almost-ordered realizations and displays fast operator growth. Inspecting the distributions, we see that for nearest-neighbor interactions much slower modes of growth are present than for dipolar interactions. These slow modes correspond to realizations where chains of next-neighbor flip-flop exchanges across distance $r$ are heavily suppressed and their extent increases with the disorder strength $h$. To explain the perceived absence of these slow modes for power-law interactions, we compare with the analytical solution $C^I_x(r,t)$ for the Ising Hamiltonian
\begin{align}\label{eq:Ising}
    \hat{H}_I = \frac{1}{2}\sum_{i,j}{\frac{C_\alpha}{r^\alpha}\siz\sjz} + \sum_i{h_i\siz}.
\end{align}
Noting that the interaction and disorder terms commute, we obtain the OTOC $\bra{\psi}\hat{U}^\dagger\six\hat{U}\sjx\hat{U}^\dagger\six\hat{U}\sjx\ket{\psi}=e^{\pm4iJ_{ij}t}$ in the computational basis with the sign being $+$($-$) for parallel (antiparallel) spins at sites $i$ and $j$. Thus, taking the infinite-temperature expectation value, the time evolution of $C^I_x(r,t)$ reads
\begin{align}\label{eq:CI}
    C^I_x(r,t) = \norm{[\six(t),\sjx]}_F^2 = 2-2\cos\left(4\frac{C_\alpha t}{r^\alpha}\right).
\end{align}
and is independent of disorder strength $h$. This function is shown in white in \cref{fig:distributions} for $\alpha=3$. When the disorder strength $h$ is sufficiently large, the nearest-neighbor model features realizations where operator growth is slower than $C_x^I(r,t)$. For power-law interactions, however, the Ising solution forms a \enquote{soft} cutoff, limiting how slow the growth can be. The disorder averages will thus only differ if $h$ is sufficiently large. Additionally, both systems feature oscillations of $C_x(r,t)$ in individual realizations.  While for the nearest-neighbor model, they wash out in the disorder average, this cutoff allows them to persist for the power-law system.

The shorter-range the power-law interactions, the higher the disorder strength $h$ must be for a significant difference between the power-law and the nearest-neighbor interacting model. This is because the cutoff is shifted to even slower modes (cf. \cref{eq:CI}). We demonstrate this in \Cref{fig:vdW} for vdW interactions ($\alpha=6$) with different disorder strengths $h=14$ and $h=21$. We recover algebraic light cones and a direct comparison shows that $C_x(r,t)$ deviates less from the corresponding nearest-neighbor case for $h=14$ than for $h=21$ and features weaker oscillations. The onset of the oscillations occurs later, as expected from the timescale $r^{-\alpha}$ in \cref{eq:CI}. In \cref{fig:vdW}~c) we plot the distribution of $C_x(r=3,t)$ at time $Jt\sim100$ across 5000 disorder realizations to explain the observed deviations by the number of realizations with growth slower than $C_x^I(r,t)$. We observe about $\sim 10$ times more such realizations for nearest-neighbor interactions than for vdW interactions and heavy-tailed or even multimodal distributions. Furthermore, the number of such realizations increases with the disorder strength $h$. We conclude that unlike ordered spin chains, where corrections are small and short-time, strongly disordered systems exhibit significantly faster operator growth and information spreading between power-law and nearest-neighbor interactions.

\section{Experimental proposal for Rydberg tweezer arrays\label{sec:exp}}
\subsection{Measurement protocol}
Rydberg atom arrays constitute a leading platform for quantum simulation, offering unprecedented control over system geometry and interactions \cite{browaeys_many-body_2020,morgado_quantum_2021,bluvstein_logical_2024}. Despite these capabilities, measuring OTOCs is notoriously difficult due to the need for echo schemes, where both $\hat{U} = e^{-i\hat{H}t}$ and $\hat{U}^\dagger = e^{i\hat{H}t}$ must be implemented. In Rydberg systems and using time-reversal, OTOCs have, to the best of our knowledge, only recently been measured in the $PXP$-model, where time-reversal was implemented on the basis of the Hamiltonian's particle-hole symmetry via a global $\sigma^z$ gate \cite{liang_observation_2025,xiang_observation_2024}, and very recently extracted via randomized global quenches \cite{toga_information_2026}. We propose a robust and general scheme to measure OTOCs in tweezer arrays with long-range interactions. We combine recently demonstrated time-reversal techniques \cite{li_tunable_2023,geier_time-reversal_2024} with Floquet engineering \cite{choi_robust_2020,geier_floquet_2021,scholl_microwave_2022,koyluoglu_floquet_2025} and the capability to locally manipulate and read out atomic states \cite{graham_multi-qubit_2022,bornet_enhancing_2024} to extract OTOCs in a broad class of Heisenberg XYZ spin models with tunable anisotropies and with or without additional on-site potentials.

\begin{figure}[!ht]
    \centering
    \includegraphics[scale=1]{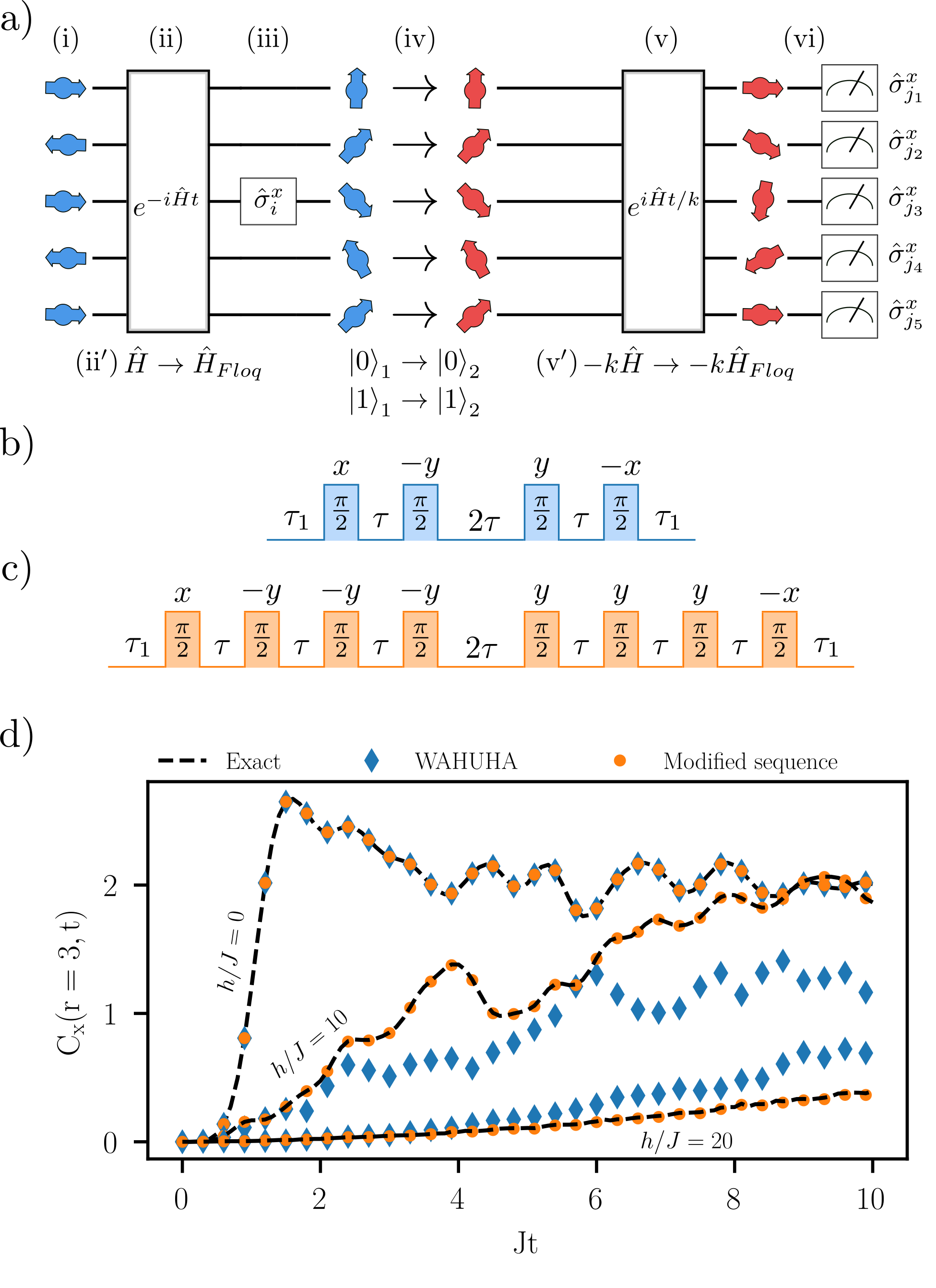}
    \caption{\textbf{Measurement Sequence} 
    \textbf{a)} Sequence of steps for measuring OTOCs. The time-reversal is achieved by transferring the states between two spin-1/2 encodings in the Rydberg manifold, effectively changing $C_3$ to $-kC_3$. Periodic driving can be applied during forward and backward evolution to realize an effective Floquet Hamiltonian.
    \textbf{b)} The WAHUHA sequence used for introducing $\siz\sjz$-interactions to obtain an XXZ model.
    \textbf{c)} Our proposed and modified sequence designed to cancel out fields in $x$- and $y$-directions and to implement the XXZ model with random on-site disorder.
    \textbf{d)} Performance of our proposed sequence. We compute the out-of-time-order commutators $C_x(r,t) = \bra{\psi}[\hat\sigma_3^x(t),\hat\sigma_j^x]^\dagger[\hat\sigma_3^x(t),\hat\sigma_j^x]\ket{\psi}$ exactly (black dashed line) and with Floquet forward and backward propagation for three disorder strengths and a cycle time of $t_c=0.1J^{-1}$. For comparison, a single, but for both sequences same, disorder realization is used. The initial state is chosen as a Néel state along $x$, the anisotropy is $\Delta=0.5$ and system size is $N=13$.}
    \label{fig:proposal}
\end{figure}

A variety of Heisenberg Hamiltonians can be implemented by microwave-coupling of different Rydberg states \cite{scholl_microwave_2022,franz_observation_2024,steinert_spatially_2023}. Our mechanism for time-reversal, i.e., changing the sign of the Hamiltonian, relies on the dipolar interaction between Rydberg $S$- and $P$-states. 
The interaction between dipolar atoms separated by $\mathbf{r}_{ij}$ is described by their electric dipole operators $\hat{\mathbf{d}}_i$ and $\hat{\mathbf{d}}_j$ and the Hamiltonian
\begin{equation}
    \hat{H}_{DD} = \frac{1}{4\pi\varepsilon_0} \sum_{i<j} \frac{\hat{\mathbf{d}}_i \cdot \hat{\mathbf{d}}_j - 3(\hat{\mathbf{d}}_i \cdot \hat{\mathbf{n}}_{ij})(\hat{\mathbf{d}}_j \cdot \hat{\mathbf{n}}_{ij})}{r_{ij}^3},
\end{equation}
where $\hat{\mathbf{n}}_{ij} = \mathbf{r}_{ij}/|\mathbf{r}_{ij}|$. The transition matrix elements yield an interaction strength $C_3(1 - 3\cos^2\theta_{ij})/r_{ij}^3$, where $\theta_{ij}$ is the angle between the internuclear separation axis and the quantization axis. Encoding $\ket{0} = \ket{nS}$ and $\ket{1} =\ket{nP}$ as (pseudo-)spins, the system is effectively described by the Heisenberg XX model
\begin{align}
    \hat{H}_{XX} = \frac{1}{2}\sum_{i,j}{J_{ij}\left(\six\sjx+\siy\sjy\right)}
\end{align}
with $J_{ij} =C_3(1-3\cos^2{\left(\theta_{ij}\right)})/r_{ij}^3$. The sign of $C_3$ is positive for $\Delta m=0$ and negative for $\Delta m = \pm1$, where $\Delta m$ is the difference in magnetic quantum numbers of the states that encode the two-level system. 

The experimental protocol for measuring, for example, a $\six\sjx$-OTOC is shown in \cref{fig:proposal}~a). In a first step (i), each spin is prepared in an eigenstate $\ket{\psi}$ of $\sjx$. After a forward evolution (ii) up to time $t$, a $\pi$-pulse $R_x(\pi)=-i\six$ is applied as a local perturbation (iii) on the atom at site $i$. Coherently transferring (iv) $\ket{nS}\rightarrow\ket{n'P'}$ and $\ket{nP}\rightarrow\ket{n''S'}$ after a forward evolution up to time $t$, the interaction coefficient can be changed to $C_3^{(2)}=-kC_3^{(1)}$ with $k=|C_3^{(2)}|/|C_3^{(1)}|$. After a backward evolution (v) up to time $t' = t/k$, the quantum state reads $\ket{\psi(t)}=e^{i\hat{H}t}\six e^{-i\hat{H}t}\ket{\psi}$ and the OTOC $\bra{\psi}\six(t)\sjx\six(t)\sjx\ket{\psi}=\pm\bra{\psi}\six(t)\sjx\six(t)\ket{\psi}$ is obtained by a measurement (vi) of $\sjx$. 

Using Hamiltonian engineering techniques, inspired by NMR tomography \cite{wei_exploring_2018, wei_emergent_2019}, we can modify this protocol to introduce $\siz\sjz$ interaction terms. To this end, we periodically drive the system with sequences of fast $\pi/2$-pulses during the forward and backward evolutions. For integer multiples of the cycle time $t=nt_c$, the dynamics is described to lowest order by the Floquet Hamiltonian
\begin{align}
    \hat{H}_{Floq} = \frac{1}{2} \sum_{i,j}{J_{ij}^\perp\left(\six\sjx+\siy\sjy\right)+J_{ij}^\parallel\siz\sjz}.
\end{align}

\Cref{fig:proposal}~b) shows the ubiquitous Waugh-Huber-Haeberlen (WAHUHA) sequence \cite{waugh_approach_1968}, previously used in \cite{geier_floquet_2021,geier_time-reversal_2024} to create XXZ and, as an extension, XYZ Hamiltonians by changing the delay between the $y$-pulses as an additional degree of freedom. Given a desired anisotropy $\Delta\in[0,2]$ and a cycle time $t_c$, applying WAHUHA with delays of $\tau_1 = \frac{t_c}{2} \frac{2-\Delta}{2+\Delta}$ and $\tau = \frac{t_c}{2} \frac{\Delta}{2+\Delta}$ results in interaction strengths $J_{ij}^\perp = J_{ij}\frac{\tau_1+\tau}{t_c}=J_{ij}\frac{2}{2+\Delta}$ and $J_{ij}^\parallel = J_{ij}\frac{2\tau}{t_c}=J_{ij}\frac{2\Delta}{2+\Delta}$. Due to the state transfer, the sign of $J_{ij}$ is changed during the backward evolution and the evolution follows $-k\hat{H}_{Floq}$. This is in contrast to Floquet experiments in NMR, where the starting point is an XXZ Hamiltonian and the Floquet sequence inverts not the sign of the interactions but the sign of the Hamiltonian. A standard WAHUHA sequence in that setting yields $-\frac{1}{5}\hat{H}_{XXZ}$, which results in an experimental duration of $6t$ needed to measure $C_x(r,t)$. Combining state transfer and Floquet engineering allows for a shorter duration of $\frac{2+\Delta}{2}(1+1/k)t$ where $k\approx1$, see \cite{geier_time-reversal_2024}. In addition, conventional Floquet time-reversal requires $\Delta = -2$, whereas our scheme unlocks $\Delta \in [0,2]$. 

On-site potentials $h_i\siz$ can be applied as light shifts from a set of off-resonant optical tweezers. The strength $h_i$ of the potential depends on tweezer intensity and detuning and can be spatially varied as demonstrated in studies on selective light shift addressing
\cite{saffman_analysis_2005,li_entanglement_2013,de_leseleuc_optical_2017,bornet_enhancing_2024}
\footnote{Technically, selective light shift addressing implements $h_i(\siz\pm1)/2$, but the second term can be neglected as it only adds a global phase to the wavefunction.}. However, the WAHUHA sequence introduces unwanted fields of strength $h_i\frac{2\Delta}{2+\Delta}$ in the $x$- and $y$-directions, known as chemical shifts \cite{choi_robust_2020}. We propose a minimal reflection-symmetric sequence, depicted in orange in \cref{fig:proposal}~c), that eliminates this effect. With  $\tau_1 = \frac{t_c}{2}\frac{2-\Delta}{2+\Delta}$ and $\tau = \frac{t_c}{4} \frac{\Delta}{2+\Delta}$, the new sequence results in the same interaction strengths as WAHUHA but in the field strengths $h_i^z = h_i\frac{2-\Delta}{2+\Delta}$ and $h_i^x=h_i^y=0$. The mechanism of that sequence can be understood by following the Hamiltonian in the toggling frame during the individual stages of the Floquet sequences. Whereas for the WAHUHA sequence the $h_i\six$-fields add $\tau_1\siz\rightarrow\tau \siy\rightarrow2\tau \six\rightarrow\tau \siy\rightarrow\tau_1 \siz$ fields, the total $\six$ and $\siy$ contributions are nulled in the modified sequence $\tau_1\siz\rightarrow\tau \siy\rightarrow\tau \six\rightarrow-\tau \siy\rightarrow-2\tau  \six\rightarrow-\tau \siy\rightarrow\tau \six\rightarrow\tau \siy\rightarrow\tau_1 \siz$, while the total ratio $J_{ij}^\parallel/J_{ij}^\perp=\Delta$ is kept identical. Because all Hamiltonians in the toggling frame respect the reflection symmetry, i.e.,~$\tilde{H}_{n+1-k}=\tilde{H}_k$ with $n$ the number of pulses, the first-order correction to $\hat{H}_{Floq}$ vanishes \cite{choi_robust_2020}, increasing  the sequence's robustness. In addition to the new driving sequence, the tweezers must be adjusted after the state transfer to ensure $h_i'=-kh_i$ as the light shift depends on laser intensity and detunings from the respectively used states. In \cref{fig:proposal}~d), we numerically show that the system driven by the modified sequence at a moderate cycle time of $t_c = 0.1J^{-1}$ faithfully reproduces the evolution of $C_x(r,t)$, whereas WAHUHA fails to follow if disorder is present.

For Rydberg atoms at distances of a few \si{\micro\meter}, interactions are on the order of a few MHz and the required cycle time of $t_c\sim100$ ns is within the capabilities of current microwave technology. Especially at large disorder ($h/J\sim10$), the tweezers inherently introduce heating via off-resonant scattering and dephasing from intensity fluctuations. Under experimental parameters similar to Refs. \cite{de_leseleuc_optical_2017,geier_time-reversal_2024}, i.e.,~$J\approx2\pi\times1$~MHz, $2\pi\times5-10$~GHz detunings, $2\pi\times 500$~MHz Rabi frequencies and $n\approx 60$ Rydberg states in Rb, the effective scattering-limited lifetime of the off-resonantly coupled $nS$ state is comparable to its blackbody-limited lifetime of $100$~\textmu s. The bare dephasing time $T_2^*\approx 16$~\textmu s at a $\delta I/I\sim10^{-3}$ intensity stabilization is comparable to the dominating $10-30$~\textmu s timescale of recapture loss from thermal atomic motion and expected to be considerably larger in practice since the echo scheme cancels out fluctuations that are slow compared to the few \textmu s experimental timescale.

\subsection{State Sampling Strategies\label{subsec:sampling}}
\begin{figure}[tbp]
    \centering
    \includegraphics[scale=1]{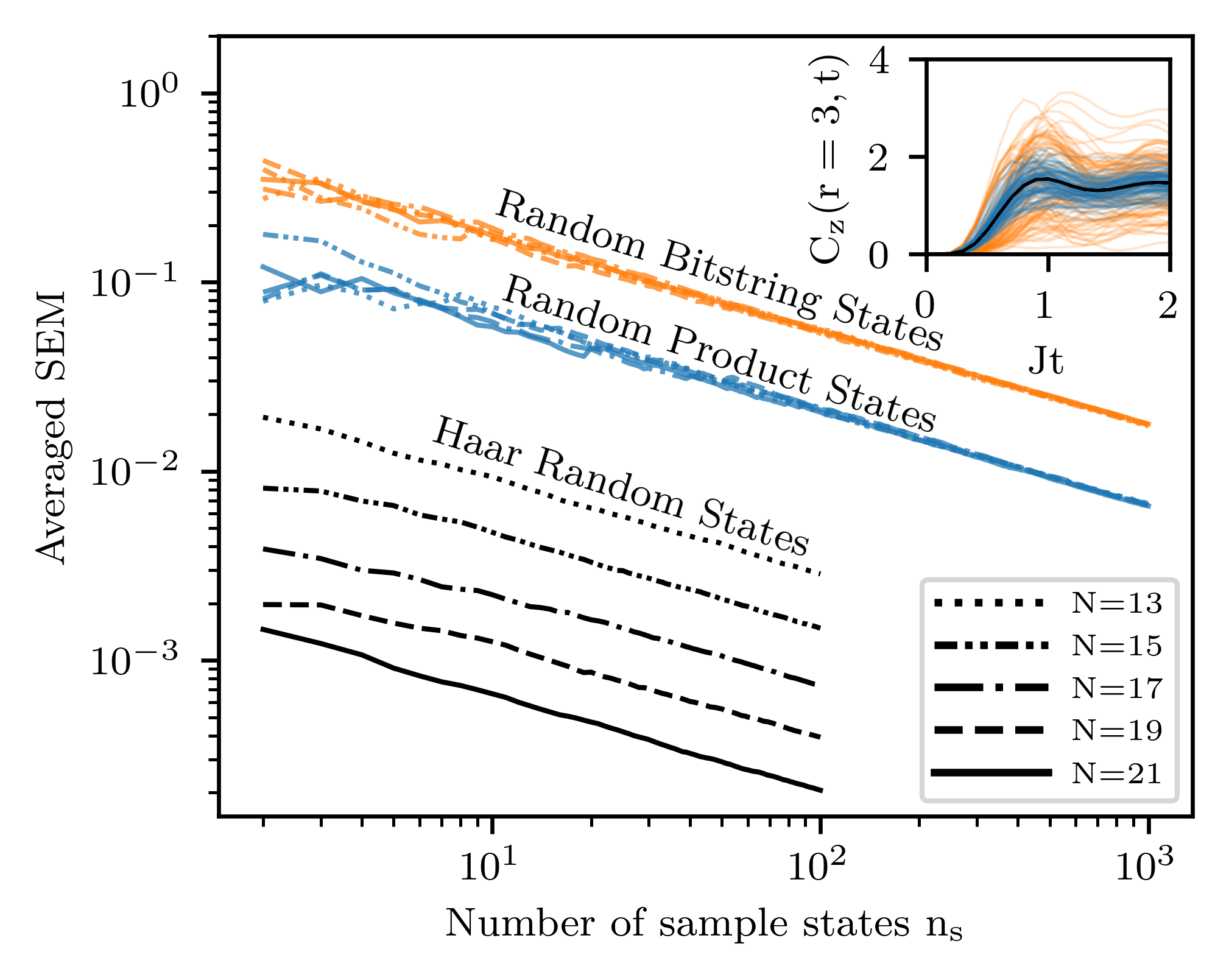}
    \caption{\textbf{Initial State Sampling} Standard error of the mean (SEM) of $\overline{C}_z$ at zero disorder $h/J=0$ as a function of number of sample states $n_s$, where $\overline{C}_z$ represents the average of $C^\psi_z(r,t)$ over all distances and 20 timesteps from $Jt = 0.1$ to $Jt = 2$. It is shown for three ensembles of initial states and system sizes $N\in\{13,15,17,19,21\}$. The inset highlights the variation of $C_z(r=3,t)$ for distinct initial states $\Ket{\psi}$ and uses the same colors for the different ensembles as the main panel.}
    \label{fig:sampling}
\end{figure}

In \cref{sec:theory}, we calculated the infinite-temperature expectation value $2^{-N}\Tr\left(\Wi(t)\Vj\Wi(t)\Vj\right)$ of the OTOC, while in \cref{sec:exp} the proposed measurements are with respect to an initial state $\ket{\psi}$. In fact, theoretical works mostly deal with the infinite-temperature state or with bounds on operator growth, whereas most experiments consider more easily producible product states. Furthermore, experiments must deal with the fact that OTOCs are not Hermitian observables. To extract OTOCs, the initial state is as above restricted to an eigenstate of $\Vj$ so that $\bra{\psi}\Wi(t)\Vj\Wi(t)\Vj\ket{\psi} = \lambda\bra{\psi}\Wi(t)\Vj\Wi(t)\ket{\psi}$ as done in \cite{garttner_measuring_2017,nie_experimental_2020,pegahan_energy-resolved_2021,zhao_probing_2022,liang_observation_2025,xiang_observation_2024}. Alternatively, a reconstruction from two separate measurements \cite{mitarai_methodology_2019,braumuller_probing_2022} or an interferometric approach \cite{swingle_measuring_2016,zhu_measurement_2016,mi_et_al_information_2021} have been used in the literature. Additionally, a protocol tailored to OTOCs of Pauli operators that circumvents initial state and ancilla requirements has recently been proposed \cite{kastner_ancilla-free_2024}. The infinite-temperature expectation value is estimated by averaging over many random initial states. In a study on disordered XX-ladder models, Dağ and Duan \cite{dag_detection_2019} asked how many Fock states one would need to sample for that. Our goal here is to determine ensembles of initial states that also minimize experimental complexity, namely the difficulty of state preparation and the number of experimental cycles necessary to reach a certain accuracy.

We compare three different ensembles of initial states. While the infinite temperature expectation value of the OTOC can be efficiently calculated with Haar random states, these states are highly-entangled and difficult to prepare experimentally. In addition, the proposed measurement scheme requires that the initial state be an eigenstate of the local operator $\Vj$. The states that satisfy this constraint and that are easy to produce are product states $\ket{\psi} = \prod_k{\ket{\psi_k}}$ where the $j$th spin is prepared as an eigenstate of $\Vj$. We will consider two possibilities here: Random bitstring states and random product states. For the former, every $\ket{\psi_k}$ is initialized in a $\hat{V}_k$ eigenstate, i.e.,~$\ket{\psi_k}$ is either $\ket{+}_x$ or $\ket{-}_x$ for a $\six\sjx$-OTOC measurement --- respectively $\ket{\uparrow}$ or $\ket{\downarrow}$ for a $\siz\sjz$-OTOC measurement as below. In contrast, the latter allows any random spin state for a spin at position $k\neq j$ \footnote{For practical purposes, we generalize this class to fully random product states. We drop the restriction that spin $j$ is in a $\hat{\sigma}_z$ eigenstate to ease averaging over distances and enable the restriction to the largest symmetry sector. Our results remain valid since the performance of such states should be worse than that of fully random product states.}.

The inset of \cref{fig:sampling} shows the $\siz\sjz$-OTO commutator $C_z(r,t) = \bra{\psi}[\siz(t),\sjz]^\dagger[\siz(t),\sjz]\ket{\psi}$ for an ordered ($h/J=0$) XXZ Heisenberg chain and the above described classes of initial states. Whereas Haar random states show little variance, $C_z$ depends greatly on the chosen initial state for the more easily preparable product states. The average over many initial states approaches the expectation value $\frac{1}{2^N}\Tr\left([\siz(t),\sjz]^\dagger[\siz(t),\sjz]\right)$, but the speed of convergence differs for different random state ensembles. Samples of random product states show less variance in $C_z$ and their average converges faster than that of random bitstring states. However, random bitstring states have the advantage that the OTOC for all sites $j$ of the spin chain can be measured simultaneously, while random product states require a series of $N$ measurements. To see which set is better at estimating the infinite-temperature expectation value, we investigate the dependence of the convergence rates on the system size $N$. We use Krylov methods \cite{park_unitary_1986,luitz_information_2017,luitz_emergent_2019,colmenarez_lieb-robinson_2020} and restrict calculations to the largest symmetry sector to increase the range of accessible system sizes compared to \cref{sec:theory}. At system sizes $N>13$, calculating the full trace of $[\siz(t),\sjz]^\dagger[\siz(t),\sjz]$ proves prohibitively expensive. As a proxy, we use the standard error of the mean (SEM), averaged over time steps from $Jt=0.1$ to $Jt=2$, shown in the main panel of \cref{fig:sampling}. Although random product states converge faster than random bitstring states, the relative error becomes independent of system size as $N$ increases. Because random bitstring states allow for simultaneous readout of $N$ observables, they prove more economical once $N\gtrsim10$. For Haar random states, we recover the exponential increase in accuracy with system size, which for unitary operators is also independent of the Hamiltonian and therefore holds equally at higher disorder (cf. \cref{app:typicality}). For random product or bitstring states, however, the sampling variance is expected to grow with disorder \cite{dag_detection_2019}.

\section{Conclusion}
Motivated by the capabilities of current quantum simulation platforms, we investigated information scrambling in strongly disordered XXZ Heisenberg chains with power-law interactions. We found algebraic light cones instead of logarithmic light cones that are well-established for the model with nearest-neighbor interactions often used in localization studies. Although in ordered one-dimensional systems, nearest-neighbor and power-law interactions lead to, up to small corrections, the same operator growth if $\alpha\geq3$ \cite{tran_lieb-robinson_2021,luitz_emergent_2019}, they significantly differ in strongly disordered systems. Intriguingly, this result holds not just for conventionally long-range dipolar interactions but also for vdW and even shorter-range interactions, which are often approximated as nearest-neighbor. Our results, together with previous studies on \enquote{algebraic many-body localization} \cite{yao_many-body_2014,pino_entanglement_2014,burin_localization_2015,de_tomasi_algebraic_2019,safavi-naini_quantum_2019,deng_universal_2020}, indicate that complex equilibrium and out-of-equilibrium phenomena induced by long-range interactions \cite{defenu_long-range_2023} can reemerge when disorder, be it on-site or positional, suppresses otherwise dominant processes. In addition, we proposed an experimental protocol that enables the extraction of OTOCs in a variety of spin models, including XXZ models with tunable anisotropy $0\leq\Delta<2$ and programmable disorder. Since the protocol relies on currently available experimental techniques such as Floquet Hamiltonian engineering and state transfer time-reversal, it can readily be applied in analog quantum simulators such as Rydberg or polar molecule tweezer experiments.

\section*{Acknowledgements}

The authors thank J. Schachenmeyer, N. Euler and M. Hornung for helpful discussions. For numerical simulations, we used the Julia programming
language \cite{bezanson_julia_2017}. M. Müllenbach acknowledges funding by the European Union H2021 MSCA Doctoral Network QLUSTER (Grant No. 101072964). E. Braun acknowledges support by the IMPRS for Quantum Dynamics in Physics, Chemistry and Biology. This research is supported by funding from the German Research Foundation (DFG) under the project identifier 398816777-SFB 1375 (NOA). The authors acknowledge support by the state of Baden-Württemberg through bwHPC and the German Research Foundation (DFG) through grant no INST 40/575-1 FUGG (JUSTUS 2 cluster). The authors acknowledge support by the state of Baden-Württemberg through bwHPC and the German Research Foundation (DFG) through grant INST 35/1597-1 FUGG (Helix cluster). This work is part of and supported by the Deutsche Forschungsgemeinschaft (DFG, German Research Foundation) under Germany’s Excellence Strategy EXC2181/1-390900948 (the Heidelberg STRUCTURES Excellence Cluster), within the Collaborative Research Centre “SFB 1225 (ISOQUANT)”, the DFG Priority Program “GiRyd 1929”, the Horizon Europe programme HORIZON-CL4-2022-QUANTUM-02-SGA via the project 101113690 (PASQuanS2.1), and the Heidelberg Center for Quantum Dynamics.

\section*{Data availability}

The data that support the findings of this study, together with the
simulation and analysis code, are openly available in Zenodo \cite{muellenbach_data_2026}.

\appendix

\section{Accuracy of Haar random state averages \label{app:typicality}}

In \cref{sec:theory,subsec:sampling} we estimate infinite-temperature expectation values by averaging over Haar random states. Here we quantify the accuracy of this estimate and show that, for the OTO commutators considered in this work, it is independent of the disorder strength. Let $\hat{A}(t)=\six(t)\sjx\six(t)\sjx$ denote the operator entering \cref{eq:xx-OTOC} and $f_t(\ket{\psi})=\bra{\psi}\hat{A}(t)\ket{\psi}$ its single-state estimate, with $\ket{\psi}$ drawn from the Haar measure on the $D=2^N$-dimensional Hilbert space. The first two moments of $f_t$ are
\begin{align}
    \mathbb{E}_\psi\left[f_t(\ket{\psi})\right]
    &= \frac{1}{D}\Tr\big(\hat{A}(t)\big)
    \label{eq:app-mean}
\end{align}
and
\begin{align}
    \mathrm{Var}_\psi\left[f_t(\ket{\psi})\right]
    &=\frac{\Tr\big(\hat{A}(t)\hat{A}^\dagger(t)\big)
        -\big|\Tr\hat{A}(t)\big|^2/D}{D(D+1)}.
    \label{eq:app-var}
\end{align}
\Cref{eq:app-mean,eq:app-var} follow from the first and second moment
of the Haar measure and hold for any quantum 2-design \cite{ullah_expectation_1963,hams_fast_2000,jin_random_2021}. We would like to stress here that the random product and random bitstring ensembles in contrast only satisfy \cref{eq:app-mean}. 

$\hat{A}(t)$ is itself unitary and thus $\Tr\big(\hat{A}(t)\hat{A}^\dagger(t)\big)=D$. The variance is then bounded as
\begin{align}
    \mathrm{Var}_\psi\left[f_t(\ket{\psi})\right]
    =\frac{1-|\langle f_t\rangle|^2}{D+1}
    \leq\frac{1}{D+1}\approx 2^{-N},
    \label{eq:app-unitary}
\end{align}
with $\langle f_t\rangle=\Tr\hat{A}(t)/D$. The bound
\cref{eq:app-unitary} holds uniformly in time and for every disorder
realization since $\Tr\hat{A}$ and $\Tr\big(\hat{A}\hat{A}^\dagger\big)$ are invariant under Heisenberg evolution. The accuracy of the estimate is therefore a kinematic property of the Haar ensemble and the Pauli algebra and does not rely on any thermalization property of the dynamics. Averaging $n_s$ independent Haar random states and using $C_x(r,t)=2(1-\mathrm{Re}\langle f_t\rangle)$, we bound the state sampling error by
\begin{align}
    \mathrm{std}\left[C_x\right]
    \leq\frac{2}{\sqrt{(D+1)\,n_s}}
    \approx\frac{2\cdot2^{-N/2}}{\sqrt{n_s}}
    \approx 7\cdot10^{-3}
    \label{eq:app-haarbound}
\end{align}
for the $N=13$ and $n_s=10$ used in \cref{sec:theory}. Although the expectation value $C_x(r,t)$ itself varies significantly across disorder realizations, the bound \cref{eq:app-haarbound} applies to each realization separately, and the statistical error of the disorder-averaged data is dominated by the disorder ensemble, not by the Haar sampling per disorder realization.

\bibliography{references}

\clearpage

\end{document}